\documentclass[aps,prd,superscriptaddress,nofootinbib,floats,preprint,floats,floatfix]{revtex4-1}
\usepackage{bm}
\usepackage{indentfirst}
\usepackage{amsmath}
\usepackage{graphicx}
\usepackage{float}
\usepackage{amssymb}
\usepackage{subfigure}
\usepackage{psfrag}
\usepackage{hyperref}
\usepackage{tensor}
\usepackage{braket}
\usepackage{multirow}
\usepackage{verbatim}
\hypersetup{
	colorlinks=true,
	linkcolor=red,
	citecolor=blue,
}
\allowdisplaybreaks[1]

 \newcommand{\bq}{\begin{equation}}
 \newcommand{\eq}{\end{equation}}
 \newcommand{\bqn}{\begin{eqnarray}}
 \newcommand{\eqn}{\end{eqnarray}}
 \newcommand{\nb}{\nonumber}
 \newcommand{\lb}{\label}

\begin{document}
\title{Scalar induced gravitational waves from Chern-Simons gravity during inflation era}

\author{Jia-Xi Feng}
\email{fengjx57@mail2.sysu.edu.cn}
\affiliation{School of Physics and Astronomy, Sun Yat-sen University, Zhuhai 519082, China}
\author{Fengge Zhang}
\email{zhangfg5@mail.sysu.edu.cn}
\affiliation{School of Physics and Astronomy, Sun Yat-sen University, Zhuhai 519082, China}
\author{Xian Gao}
\email{Corresponding author: gaoxian@mail.sysu.edu.cn}
\affiliation{School of Physics and Astronomy, Sun Yat-sen University, Zhuhai 519082, China}

\begin{abstract}
We investigate the scalar induced gravitational waves (SIGWs) in the Chern-Simons (CS) gravity with a dynamical scalar field during slow roll inflation. Due to the parity violation in the CS term, the SIGWs are generally polarized, which are effectively characterized by the degree of circular polarization. We derive the semianalytic expression to evaluate the power spectra and the degree of circular polarization of the SIGWs, which receive contributions from the general relativity and the parity-violating term, respectively. We find that the correction from the parity-violating CS term is negligible on large scales, which means that the degree of circular polarization of SIGWs is very small.
\end{abstract}

\maketitle

\vspace*{0.5cm}

\section{Introduction}\label{s1}
The detection of the gravitational waves (GWs)  by
LIGO/Virgo \cite{LIGOScientific:2016fpe,LIGOScientific:2016sjg,LIGOScientific:2016aoc,LIGOScientific:2017vox,LIGOScientific:2017vwq,LIGOScientific:2017ycc,LIGOScientific:2017bnn,LIGOScientific:2018mvr,LIGOScientific:2020zkf,LIGOScientific:2020aai,LIGOScientific:2020stg}
opens a new window to explore the nature of gravity.  In addition to the GWs  from astrophysical sources, there are also various GW signals from cosmological origins 
\cite{Caprini:2018mtu}, which compose the stochastic gravitational waves background (SGWB).
The so-called scalar induced gravitational waves (SIGWs) sourced by the first-order cosmological perturbations \cite{Kodama:1984ziu,Mukhanov:1990me,Ananda:2006af,Baumann:2007zm}, are an important part of SGWB. Recently, SIGWs have attracted much attention, and there are also a large number of works on exploring the early universe and cosmological perturbations on small scales with SIGWs \cite{Saito:2008jc,Orlofsky:2016vbd,Nakama:2016gzw,Wang:2016ana,Cai:2018dig,Kohri:2018awv,Espinosa:2018eve,Kuroyanagi:2018csn,Domenech:2019quo,Fumagalli:2020nvq,Lin:2020goi,Domenech:2020kqm,Domenech:2021ztg,Zhang:2021vak,Wang:2021djr,Adshead:2021hnm,Garcia-Saenz:2022tzu,Ahmed:2021ucx,Zhang:2021rqs,Zhou:2021vcw,Solbi:2021rse,Zhang:2022xmm,Romero-Rodriguez:2021aws,Cai:2021wzd,Chen:2021nxo,Kozaczuk:2021wcl,Inomata:2021zel,Rezazadeh:2021clf,Davies:2021loj,Witkowski:2021raz,Ota:2021fdv,Yi:2022ymw,Papanikolaou:2021uhe,Balaji:2022rsy,Arya:2022xzc,Yi:2022anu,Aldabergenov:2022rfc,Zhang:2022dgx,Fu:2022ssq}.
Especially, there is evidence that the signal of SIGWs  may have been detected according to North American Nanohertz Observatory for Gravitational Waves (NANOGrav) 12.5 yrs data \cite{DeLuca:2020agl,Vaskonen:2020lbd,Kohri:2020qqd,Domenech:2020ers}. Moreover, cosmological GWs may be detected by future GW detectors, such as LISA \cite{Danzmann:1997hm,LISA:2017pwj}, DECIGO \cite{Kawamura:2020pcg}, Taiji \cite{Hu:2017mde} and TianQin \cite{TianQin:2015yph} and the Pulsar Timing Array (PTA) \cite{Kramer:2013kea,Hobbs:2009yy,McLaughlin:2013ira,Hobbs:2013aka}, etc.

Considering gravity theories beyond general relativity (GR), especially the gravity theories with parity-violating (PV) terms have attracted more and more attention in recent years \cite{Horava:2009uw,Crisostomi:2017ugk,Gao:2019liu,Hu:2021bbo,Hu:2021yaq,Zhu:2022uoq,Odintsov:2021kup,Nojiri:2019nar,Nojiri:2020pqr,Odintsov:2020iui,Odintsov:2022cbm}. Based on the Riemannian geometry, the simplest PV term we can construct is the Chern-Simons (CS) term.
It was first proposed by embedding the three-dimensional CS term into the four-dimensional GR \cite{Jackiw:2003pm}.
Since then, the GWs and cosmology have been studied extensively in the CS gravity \cite{Lue:1998mq,Satoh:2007gn,Saito:2007kt,Satoh:2007gn,Alexander:2009tp,Yunes:2010yf,Gluscevic:2010vv,Myung:2014jha,Kawai:2017kqt,Nishizawa:2018srh,Nair:2019iur,Odintsov:2019mlf,Zhao:2019xmm,Li:2022grj,Qiao:2022mln} and other parity-violating gravity theories, such as the Nieh-Yan term \cite{Nieh:1981ww,Chatzistavrakidis:2020wum,Langvik:2020nrs,Li:2020xjt,Cai:2021uup,Wu:2021ndf,Li:2021wij,Rao:2021azn,Li:2021mdp,Li:2022mti}, as well as in more general models with non-vanishing torsion and/or non-metricity tensors  \cite{Hohmann:2020dgy,Bombacigno:2021bpk,Iosifidis:2020dck,Hohmann:2022wrk,Conroy:2019ibo,Iosifidis:2021bad,Pagani:2015ema,Li:2022vtn}.
Interestingly, the GWs in PV gravity show diverse features that are different from GR, such as the velocity and amplitude birefringence phenomena of GWs \cite{Takahashi:2009wc,Wang:2012fi,Alexander:2004wk,Wang:2020pgu,Wang:2021gqm,Wang:2020cub,Zhao:2022pun}.
Furthermore, the GWs generated during inflation will produce non-vanishing TB and EB spectra on the cosmic microwave background (CMB) besides TT, EE, BB, and TE spectra due to parity violation \cite{Lue:1998mq,Saito:2007kt,Seto:2007tn,Qiao:2019hkz}. The observations on CMB may provide some evidence to test parity violation in the early universe.

All the works on the GWs in PV gravity theories mentioned above mainly concentrated on the primordial or linear GWs. In this work, we shall pay attention to the SIGWs in PV gravity. In our previous work \cite{Zhang:2022xmm}, we used the SIGWs to test  PV phenomena for the first time, in which we focus on  CS gravity coupled with a dynamical scalar field during the radiation-dominated era. We find that the degree of circular polarization can be quite large. In order to investigate the possible parity violation of gravity interaction in the early universe, in this paper, we consider the SIGWs generated during slow roll inflation in detail.
For simplicity of discussion, we consider the coupling function to be an exponential form, then we obtain a semianalytic expression for the power spectra $\mathcal{P}^A_{h}$ of SIGWs. We also briefly discuss the case where the coupling function is linear.

The rest of the paper is organized as follows. In section \ref{s2}, we briefly  review the CS gravity and equations of motion for the SIGWs. In section \ref{s3}, we show the basic formulas related to SIGWs during slow roll inflation. We derive the semianalytic expressions for the power spectra of SIGWs in section \ref{s4}.
Our main conclusions are summarized in section \ref{s5}.

There are also two appendices, \ref{app1} and \ref{app2}, in which some equations and calculations involved in this paper are given in detail.

\section{Scalar Induced Gravitational Waves from Chern-Simons gravity  }\label{s2}

In this section, we first briefly introduce  CS gravity, which is the most popularly studied example of parity-violating gravity. Then we list the equations of motion (EOM)  of SIGWs. The EOM for the background evolution and the first-order scalar perturbation can be found in appendix \ref{app1}.

\subsection{Chern-Simons gravity}

We consider Chern-Simons gravity, of which the action is of the form
\begin{equation}
\label{action}
S= \frac{1}{16\pi G}\int \mathrm{d}^{4} x~ \sqrt{-g}\Big(R+\mathcal{L}_{\mathrm{CS}}\Big)+\int \mathrm{d}^{4} x~ \sqrt{-g}~\mathcal{L}_{\varphi},
\end{equation}
where
\begin{equation}
\mathcal{L}_{\mathrm{CS}}=\frac18 f(\varphi)\varepsilon^{\mu \nu \rho \sigma} R_{\rho \sigma \alpha \beta} R^{\alpha \beta}{}{}_{\mu \nu},\quad\quad \mathcal{L}_{\varphi}=-\frac{1}{2}g^{\mu\nu}(\nabla_{\mu}\varphi )( \nabla_{\nu} \varphi)-V(\varphi),
\end{equation}
here
$\varepsilon^{\mu \nu \rho \sigma}$  is the Levi-Civita tensor defined by $\varepsilon^{\rho \sigma \alpha \beta}=\epsilon^{\rho \sigma \alpha \beta}/\sqrt{-g}$.\footnote{Note that $\epsilon^{\rho \sigma \alpha \beta}$ is the Levi-Civita symbol with $\epsilon_{0123}= -\epsilon^{0123} =1$.}

By varying the action \eqref{action} with respect to the spacetime metric $g_{\mu\nu}$, we obtain the following equation of motion
\begin{equation}
\label{EE}
G_{\mu \nu}+C_{\mu \nu}=\kappa^{2}T_{\mu \nu}^{\varphi},
\end{equation}
where  we set $\kappa^{2}=8\pi G$, $C^{\mu \nu}$ is the four-dimensional Cotton tensor \cite{Alexander:2009tp}
\begin{equation}
 C_{\mu \nu}=\frac12\left[v_{\alpha}\left(\varepsilon^{\alpha\beta\lambda}{}_{\mu}\nabla_{\lambda} R_{\nu\beta}+\varepsilon^{\alpha\beta\lambda}{}_{\nu}\nabla_{\lambda} R_{\mu\beta}\right)\right.
\left.+v_{\alpha\beta}
 \left({}^{*}R^{\beta}{}_{\mu\nu}{}^{\alpha}+{}^{*} R^{\beta}{}_{\nu \mu}{}^{\alpha}\right)\right],
\end{equation}
 with
 $$v_{\alpha}=\nabla_{\alpha} f(\varphi),\quad   v_{\alpha \beta}=\nabla_{\alpha} \nabla_{\beta} f(\varphi)=\nabla_{(\alpha} \nabla_{\beta)} f(\varphi), \quad  {}^{*} R^{\beta}{}_{\mu\nu}{}^{\alpha}=\frac{1}{2} \varepsilon_{\nu}{}^{\alpha\lambda\sigma} R^{\beta}{}_{\mu\lambda\sigma},$$
and the  energy-momentum tensor of the scalar field  $T_{\mu \nu}^{\varphi}$ is given by
\begin{equation}
T_{\mu \nu}^{\varphi}=\nabla_{\mu} \varphi \nabla_{\nu} \varphi- g_{\mu \nu}\left(\frac{1}{2}\nabla_{\alpha} \varphi \nabla^{\alpha} \varphi+ V(\varphi)\right).
\end{equation}
Note that according to the Hamiltonian analysis \cite{Crisostomi:2017ugk}, there are 5 degrees of freedom (DOFs) in dynamical Chern-Simons modified gravity, of which 2 or 3 DOFs are Ostrogradsky modes due to the presence of higher-order time derivatives. However, as being suggested  in  \cite{Crisostomi:2017ugk}, these modes could be ignored if the Chern-Simons term is considered as a perturbative correction to GR within the framework of Effective Field Theory (EFT). This is further supported by the absence of higher-order time derivative terms in the EOM for SIGWs, as discussed in the subsequent section.

Varying action \eqref{action}  with respect to scalar field $\varphi$,   we have
\begin{equation}
\label{seq}
 \kappa^{2}(\Box\varphi-V_\varphi)+ \frac18f_\varphi\cdot\varepsilon^{\mu \nu \rho \sigma} R_{\rho \sigma \alpha \beta} R^{\alpha \beta}{}_{\mu \nu}=0,
\end{equation}
where $\Box=\nabla^{\mu}\nabla_{\mu}=g^{\mu\nu}\nabla_{\mu}\nabla_{\nu}$, $V_\varphi=\mathrm{d}V/\mathrm{d}\varphi$, and $f_\varphi=\mathrm{d}f/\mathrm{d}\varphi$.

\subsection{The SIGWs}\label{II2}

 The perturbed metric is
\begin{equation}
\mathrm{d} s^{2}=g_{\mu \nu} \mathrm{d} x^{\mu} \mathrm{d} x^{\nu}=-a^{2}(1+2 \phi) \mathrm{d} \eta^{2}+a^{2}\left((1-2  \psi) \delta_{i j}+ \frac12h_{i j}\right) \mathrm{d} x^{i} \mathrm{~d} x^{j},
\end{equation}
where we choose the longitudinal (conformal Newtonian) gauge and neglect the anisotropic stress \footnote{In this paper, we consider the case that the universe is filled with a single scalar field with a canonical kinetic term during inflation era, of which the energy-momentum tensor has the form of a perfect fluid, so the anisotropic stress vanishes \cite{Malik:2008im}.}, and it follows that $\phi=\psi$. Here, $\phi$, $\psi$ are the first-order scalar perturbation  and $h_{i j}$ is the second-order tensor perturbation. And we do not consider the vector perturbations and the first-order GWs.

Recently, a subtle issue of SIGWs is the gauge dependence of tensor perturbations at nonlinear orders, which has been studied in \cite{Chang:2020iji,Domenech:2020xin,Lu:2020diy,Ali:2020sfw,Lin:2021vwc,DeLuca:2019ufz}.
To the best of our knowledge, although there is no universally accepted solution to this issue, it has been argued that it is appropriate to calculate SIGWs in the Newtonian gauge and will obtain gauge invariant results, at last in the radiation/matter domination era. A thorough treatment of the gauge issue is out of the scope of our current paper, which we will leave for future discussion.

In order to compute the SIGWs, we expand the $i-j$ components of eq. \eqref{EE} to the second order,
\begin{equation}
\label{G2}
G_{ij}^{(2)}+C^{(2)}_{ij}=\kappa^{2}T_{ij}^{(2)},
\end{equation}
where
\begin{align}
G^{(2)}_{ij}=&\frac14h''_{ij}+\frac12\mathcal{H}h'_{ij}-\frac12(\mathcal{H}^2+2\mathcal{H}') h_{ij}-\frac14\nabla^2{h_{ij}}+2(\partial_i\psi)(\partial_j\psi)\nb\\
&~~+4\psi(\partial_j\partial_i\psi) + (\mathrm{diagonal~ part})\delta_{ij},\\
&~~~~~~~~~~~~~~~~~\nb\\
C^{(2)}_{ij}=&-\Big[\frac{{\epsilon}_{ilk}}{a^2}\Big(f_\varphi \partial^k \delta \varphi' \partial^l\partial_j\psi+f_\varphi \partial^k \delta \varphi \partial^l\partial_j\psi'+f_{\varphi\varphi}\varphi' \partial^k \delta \varphi \partial^l\partial_j\psi\Big)+i\leftrightarrow j\Big]\nb\\
&~~+\Big[\frac{{\epsilon}_{ilk}}{8a^2}\left(f''\partial^l h_{j}^{'k}+f'\partial^{l}h_{j}^{'' k}-f'\partial^{l}\nabla^2h_{j}^{\ k}\right)+i\leftrightarrow j\Big],
\end{align}
and
\begin{equation}
T^{(2)}_{ij}=\partial_{i}\delta\varphi\partial_{j}\delta\varphi-\frac{1}{2\kappa^2}(\mathcal{H}^2+2\mathcal{H}') h_{ij}+ (\mathrm{diagonal~ part})\delta_{ij},
\end{equation}
here the prime represents the derivation with respect to the conformal time $\eta$, $\delta\varphi$ represents the fluctuation of the scalar field $\varphi$, and $f_{\varphi\varphi}=\mathrm d^2f/\mathrm d\varphi^2$.

Extracting the transverse, traceless part of eq. \eqref{G2}, we obtain
\begin{equation}
\label{G1}
{h^{}_{ij}}''+2\mathcal{H}{h^{}_{ij}}'-\nabla^2{h^{}_{ij}}+ \left[\frac{{\epsilon}_{ilk}}{2a^2}\left(f''\partial^l h_{j}^{'k}+f'\partial^{l}h_{j}^{'' k}-f'\partial^{l}\nabla^2h_{j}^{\ k}\right)+i\leftrightarrow j\right] = 4~\hat{\mathcal{T}}_{ij}^{lm}\mathcal{S}_{lm},
\end{equation}
where $\hat{\mathcal{T}}_{ij}^{lm}$, of which the definition is given below, is the projection operator extracting the traceless-transverse part of any tensor. The source term $\mathcal{S}_{ij}$ can be split into two parts,
\begin{equation}
\mathcal{S}_{ij}= {\mathcal{S}_{ij}}^{(\mathrm{GR})}+\mathcal{S}_{ij}^{(\mathrm{PV})},
\end{equation}
with
\begin{align}
{\mathcal{S}_{ij}}^{(\mathrm{GR})}&=-2(\partial_i\psi)(\partial_j\psi)-4\psi(\partial_j\partial_i\psi)+\kappa^2\partial_{i}\delta\varphi\partial_{j}\delta\varphi,
\end{align}
which is the standard contribution from the GR, and
\begin{align}
{\mathcal{S}_{ij}}^{(\mathrm{PV})}&=\frac{1}{a^2}\left[{\epsilon}_{ilk}\left(f_\varphi \partial^k \delta \varphi' \partial^l\partial_j\psi+f_\varphi \partial^k \delta \varphi \partial^l\partial_j\psi'+f_{\varphi\varphi}\varphi' \partial^k \delta \varphi \partial^l\partial_j\psi\right)+i\leftrightarrow j\right],
\end{align}
which is the correction due to the parity-violating CS term. From  the EOM of SIGWs \eqref{G1} and the EOMs for the linear scalar perturbations eqs. (\ref{eom100})-(\ref{seq1}), it is clear that there are no higher-order time derivative terms. This implies that the Ostrogradsky modes can be neglected in our  discussion.

The second-order tensor perturbation $h_{ij}$ can be decomposed into the circularly polarized modes as follows
\begin{equation}
h_{i j}(\bm x,\eta) =\sum_{A=R,L}\int \frac{\mathrm{d}^{3} \bm{k}}{(2 \pi)^{3}}e^{\mathrm{i} \bm{k} \cdot \bm x} \mathbf e_{i j}^{A}(\bm{k}) h_{\bm k}^{A}(\eta),
\end{equation}
where
$\mathbf e_{i j}^{A}$ are the circular polarization tensors defined by
\begin{equation}
\mathbf e_{i j}^{R}=\frac{1}{\sqrt{2}}(\mathbf e_{i j}^{+}+\mathrm{i}~ \mathbf e_{i j}^{\times}), \quad
\mathbf e_{i j}^{L}= \frac{1}{\sqrt{2}}(\mathbf e_{i j}^{+}-\mathrm{i}~\mathbf e_{i j}^{\times}),
\end{equation}
with
\begin{equation}
\mathbf e^+_{ij}=\frac{1}{{\sqrt{2}}}(\mathbf e_i \mathbf e_j-\bar{\mathbf e}_i \bar{\mathbf e}_j), \quad
\mathbf e_{ij}^\times=\frac{1}{{\sqrt{2}}}(\mathbf e_i\bar{\mathbf e}_j+\bar{\mathbf e}_i \mathbf e_j).
\end{equation}
In the above, ${\mathbf e_{i}\left(\bm{k}\right)}$ and ${\bar{\mathbf e}_{i}\left(\bm{k}\right)}$ are two basis vectors which are orthogonal to each other and perpendicular to the wave vector ${\bm{k}}$, i.e., satisfying ${\bm k}\cdot {\mathbf e}={\bm k}\cdot \bar{\mathbf e}={\mathbf e}\cdot \bar{\mathbf e}=0$ and $|{\mathbf e}|=|\bar{\mathbf e}|=1$.

The definition of the projection tensor is
\begin{equation}
\mathcal{T}^{lm}_{\ \ \ ij}\mathcal{S}_{lm}(\bm{x},\eta)=\sum\limits_{A=R,L}\int \frac{\mathrm{d}^3\bm k}{(2\pi)^{3}}e^{\mathrm{i} {\bm k} \cdot {\bm x}}\mathbf e_{ij}^A \mathbf e^{Alm}\tilde{\mathcal{S}}_{lm}(\bm k,\eta),
\end{equation}
where $\tilde{\mathcal{S}}_{ij}$ is the Fourier transformation of the source $\mathcal{S}_{ij}$.

With the above calculation, eq. \eqref{G1} in Fourier space becomes
\begin{equation}
\lb{TeomV}
v_{\bm k}^{A''}+\Big(k^2-\frac{B^{A''}}{B^A}\Big)v_{\bm k}^{A}=\frac{4B^A}{z^{A}}  \mathcal S_{\bm k}^{A},
\end{equation}
where $v^A_{\bm k} =B^Ah^A_{\bm k}$, and
\begin{equation}
\lb{B}
B^A(k,\eta)=a(\eta)\sqrt{z^A(k,\eta)},
\end{equation}
with
\begin{equation}
\label{zA}
    z^{A}(k,\eta)=1-\frac{k\lambda^A f'}{a^2(\eta)},\ \ \  (\lambda^{\mathrm{R}}=+1,~~ \lambda^{\mathrm{L}}=-1).
\end{equation}
The source term on the right-hand side of eq. \eqref{TeomV}
\begin{equation}
\mathcal S^A_{\bm k}=\mathbf e^{Alm}\tilde{\mathcal{S}}_{lm}(\bm k)= \mathcal S^{A(\mathrm{GR})}_{\bm k}+\mathcal S^{A(\mathrm{PV})}_{\bm k},
\end{equation}
where
\begin{equation}
\mathcal S^{A(\mathrm{GR})}_{\bm k}= \int \frac{\mathrm{d}^3p}{(2\pi)^{3}}\mathbf e_{ij}^{A}~p^i p^j \left[\kappa^2\delta\varphi_{\boldsymbol{p}}\delta\varphi_{\boldsymbol{k}-\boldsymbol{p}}+2\left(\psi_{\boldsymbol{p}}\psi_{\boldsymbol{k}-\boldsymbol{p}}\right)\right],
\end{equation}
\begin{align}
\mathcal  S^{A(\mathrm{PV})}_{\bm k}&=\int \frac{\mathrm{d}^{3} p}{(2 \pi)^{3}} \mathbf{e}_{i j}^{A}~ p^{i} p^{j}\left(-\frac{\lambda^A k}{a^2} \right)
\Big[\Big(f_{\varphi}\delta\varphi'_{\boldsymbol{p}}\psi_{\boldsymbol{k}-\boldsymbol{p}}+f_{\varphi}\delta\varphi_{\boldsymbol{p}}\psi'_{\boldsymbol{k}-\boldsymbol{p}}+f_{\varphi\varphi}\varphi'\delta\varphi_{\boldsymbol{p}}\psi_{\boldsymbol{k}-\boldsymbol{p}}\Big)\nb\\
&~~~~+\left({\boldsymbol{p}}\leftrightarrow {\boldsymbol{k}-\boldsymbol{p}} \right)\Big]. \label{calSA}
\end{align}
In the above, we have used the relation $k_l\epsilon^{i l k} \mathbf e_{j k}^{A}=-\mathrm{i} k\lambda^{A}( \mathbf e^i_{j})^{ A}$.

According to eq. \eqref{zA}, in the standard case of GR (i.e., without the CS term), $z^{A} = 1$ and thus $B^{A} = a$, and eq. \eqref{TeomV} reduces to the familiar form in the GR. In our case, however,
\begin{equation}
z^A=1-\frac{\lambda^A k_{phys}}{M_{\mathrm{PV}}},
\end{equation}
where $k_{phys}=k/a$ is the physical wavenumber and $M_{\mathrm{PV}}=a/|f'|$ is the parity-violating energy scale. On one hand, without loss of generality we assume that $f'>0$, then the factor $z^A$ in eq. \eqref{TeomV}  may become negative if $k_{phys}/M_{\mathrm{PV}}>1$ for the right-hand polarized mode. On the other hand, from the point of view of the action, the coefficient of the kinetic term of GWs in the action of GWs is proportional to $(B^A)^2=a^2z^A$ \cite{Bartolo:2017szm,Bartolo:2018elp}. It is obvious that one of the polarized modes of SIGWs will be a ghost field if $k_{\text{phys}}>M_{\rm{PV}}$. To avoid these problems, we require that
\begin{equation}
\label{kphys}
\frac{k_{phys}}{M_{\mathrm{PV}}}<1,
\end{equation}
so CS gravity is viewed only as a low-energy effective field theory.

The eq. \eqref{TeomV} can be solved by the method of Green's function,
\begin{equation}\label{Eh}
h^{A}_{\bm k}\left(\eta\right)=\frac{4}{B^A(k,\eta)}\int^{\eta}_{\eta_i}\mathrm{d}\bar{\eta}~G^A_{k}\left(\eta,\bar{\eta}\right)
\frac{B^A(k,\bar{\eta})}{z^A(k,\bar{\eta})}\mathcal S^A_{\bm k}\left(\bar{\eta}\right),
\end{equation}
where $\eta_i$ represents the beginning of inflation and the Green's function $G^A_{k}\left(\eta,\bar{\eta}\right)$ satisfies
\begin{equation}\label{GREEN}
G^{A''}_{\bm{k}}(\eta,\bar{\eta})+\Big(k^2-\frac{B^{A''}}{B^A}\Big)G^{A}_{\bm{k}}(\eta,\bar{\eta})=\delta(\eta-\bar{\eta}).
\end{equation}

From the above eq. \eqref{Eh}, in order to calculate the SIGWs, it is necessary to determine how the perturbations $\delta \varphi_{\bm k}$ and $\psi_{\bm k}$ evolve and the Green's function $G^A_{\bm k}\left(\eta,\bar{\eta}\right)$, namely, the solution to eq. \eqref{GREEN}. This is what we will do in the next section.

\section{Scalar induced Gravitational Waves during slow roll inflation}\label{s3}

Inflation is a period before the hot big bang, during which the universe undergoes an epoch of accelerated expansion. A scalar field whose potential energy dominated over the kinetic energy may give rise to a period of inflation.
In this section, let us focus on the SIGWs during slow roll inflation driven by $\varphi$, which is the scalar field CS term coupled to.

The cosmological perturbations originate from quantum fluctuations during inflation. The perturbations $\delta\varphi_{\bm k}$ and $\psi_{\bm k}$ can be expanded as
\begin{equation}
\delta\varphi_{\bm k}=U_{\delta\varphi}(k,\eta)\hat a(\bm k)+U_{\delta\varphi}^{\ast}(k,\eta)\hat a^{\dagger}(-\bm k),
\end{equation}
\begin{equation}
\psi_{\bm k}=U_{\psi}(k,\eta)\hat a(\bm k)+U_{\psi}^{\ast}(k,\eta)\hat a^{\dagger}(-\bm k),
\end{equation}
with operators $\hat a^{\dagger}(\bm k)$ and $\hat a(\bm k)$ are the creation and annihilation operators, respectively, which satisfy the standard canonical commutation relation. Note according to eq. \eqref{eom10i}, $\delta\varphi$ is determined by $\psi$, and thus they share the same set of creation/annihilation operators.
An asterisk ``$\ast$'' represents complex conjugation. The mode functions $U_{\delta\varphi}$ and $U_{\psi}$ obey the following equations according to eqs. \eqref{eom10i} and \eqref{seq1}
\begin{equation}\label{f1}
U_{\psi}'+\mathcal{H}U_\psi=\epsilon \mathcal{H}^2\frac{U_{\delta \varphi}}{\varphi'},
\end{equation}
\begin{equation}\label{f2}
U^{''}_{\delta\varphi}+2\mathcal{H}U^{'}_{\delta \varphi}+k^2 U_{\delta \varphi}-4 U^{'}_{\psi} \varphi'=-a^{2}U_{\delta\varphi} V_{\varphi\varphi}-2U_{\psi}V_{\varphi},
\end{equation}
where $\epsilon$ is the slow roll parameter, which is defined by
\begin{equation}
\label{sr}
\epsilon=-\frac{\dot H}{H^2}=\frac{\kappa^2}{2}\left(\frac{\varphi^{\prime 2}}{\mathcal{H}^2}\right).
\end{equation}

We assume that the universe experienced a near-exponential expansion, which implies that $\epsilon \ll1$ and is approximately constant during slow roll inflation.
As a result, $H$, the Hubble parameter is also approximately constant and $\mathcal{H}=a^{\prime} / a\simeq-1/\eta$ with $a\simeq-1/(\eta H)$. With the above approximation, from  eq. \eqref{sr} we obtain
\begin{equation}\label{vphip}
\varphi'= \pm\sqrt{2\epsilon}\mathcal{H}/\kappa,
\end{equation}
with the solution
\begin{equation}\label{vphi}
\varphi\simeq\frac{\sqrt{2\epsilon}\beta}{\kappa} \ln\left(\frac{\eta}{\eta_0}\right)+\varphi_0,
\end{equation}
where $\varphi_0$ is an integral constant corresponding to the value of $\varphi$ at some pivot time $\eta_0$. Here for convenience, we use $\beta$ to represent $\pm 1$.

Combing eqs. \eqref{f1} and \eqref{f2} we can derive an equation of motion for the single mode function $U_{\delta\varphi}$. This procedure can be simplified by noting that $U_{\psi} \sim \sqrt{\epsilon} U_{\delta\varphi}$, which is suppressed by the slow roll parameter. Hence we can safely neglect the contribution of $\psi$ in eq. \eqref{f2}. Moreover, a nearly flat potential is required to realize the slow roll inflation, the first term on the right-hand side of eq. \eqref{f2}, which corresponds to the mass term of inflaton, is negligible. Then we can easily obtain the approximate solution for $U_{\delta\varphi}$ from eq. \eqref{f2},
\begin{equation}
\label{U-varphi}
U_{\delta\varphi}(k,\eta)=\frac{\mathrm{i}H}{\sqrt{2k^3}}(1+\mathrm{i}k\eta )\mathrm{e} ^{-\mathrm{i}k\eta},
\end{equation}
which corresponds to the Bunch-Davies vacuum \cite{Bunch:1978yq,Chernikov:1968zm}. eq. \eqref{U-varphi} has  the same  form as that in \cite{Chen:2006nt,Wang:2013zva}.
Substituting the mode function \eqref{U-varphi} into eq. \eqref{f1}, we obtain the mode function for $\psi_{\bm k}$
\begin{equation}
\lb{U-psi}
U_{\psi}(k,\eta)=\frac{\sqrt{2\epsilon}\kappa\beta }{2}\frac{\mathrm{i}H}{\sqrt{2k^3}}\mathrm{e} ^{-\mathrm{i}k\eta}+c_1 \eta,
\end{equation}
where $c_1$ is an integral constant, which is fixed to be $0$ due to the finiteness of the mode function $U_{\psi}$ at the beginning of inflation, $\eta\rightarrow-\infty$.

In order to obtain the analytic expression for the Green's function, we need to determine the solution for the homogeneous equations corresponding to eq. \eqref{TeomV}. To this end, first we take the exponential form of the coupling function \cite{Zhang:2022xmm}
\begin{equation}
\label{assume}
f(\varphi)=f_0 e^{\kappa\alpha\varphi},
\end{equation}
where $f_0$ and $\alpha$ are constants. With the expressions of $\varphi'$  \eqref{vphip} and $\varphi$ \eqref{vphi},  $z^A$ in eq. \eqref{zA} becomes
\begin{equation}
z^A(k,\eta)=1-\frac{k\lambda^A  f_0H^2(\sqrt{2\epsilon}\alpha\beta) e^{\kappa\alpha\varphi_0}}{\eta_0^{ \sqrt{2\epsilon}\alpha\beta }}\eta^{ \sqrt{2\epsilon}\alpha\beta + 1}.
\end{equation}
Clearly, due to the  parity-violating nature of the CS gravity, the solutions for the homogeneous equations of motion are different from those in GR, which will result in a different Green's function from the case of GR. Nevertheless, in this work we focus on the correction due to the PV source term $\mathcal S^{A(\mathrm{PV})}$ in eq. \eqref{calSA}. Therefore we choose the simplest expressions for the Green's function, as we will do below. This is also justified by the fact that correction due to the modification of the Green's function will be sub-leading order compared with the correction due to the PV source term.

If $z^A$ is independent of $\eta$, then we can easily solve the eq. \eqref{GREEN} to obtain the analytic expression for the Green's function, which is the same as that in GR, coincidently.
Since in this work we focus on the effect due to the parity-violating CS term, we assume that the constant value of $\alpha$ is chosen such that $\sqrt{2\epsilon}\alpha\beta+1=0$, and thus
\begin{equation}
\label{coz}
z^A(k)=1-z_0k\lambda^A,
\end{equation}
where $z_0=-f_0H^2\eta_0 e^{\kappa\alpha\varphi_0}$.
As we mentioned in section \ref{s2}, $z_0 k=k_{phys}/M_{\mathrm{PV}}< 1$  should be satisfied to avoid the ghost field. Under this assumption, the Green's function is
\begin{equation}\label{GGreen}
G^A_{\bm k}(\eta,\bar{\eta})=\Theta\left(\eta-\bar{\eta}\right) \frac{1}{k^{3}  \eta\bar{\eta}
}\Big\{k\left(\bar{\eta}-\eta\right) \cos\Big[k\left(\bar{\eta}-\eta\right)\Big]-\left(1+k^{2} \eta \bar{\eta}\right) \sin \Big[k\left(\bar{\eta}-\eta\right)\Big]\Big\}.
\end{equation}

With the above expressions for the perturbations $\delta\varphi$, $\psi$ and the Green's function $G^A_{\bm k}(\eta,\bar{\eta})$, we can compute the SIGWs and their power spectra.

\section{Power spectra and the degree of circular polarization of SIGWs}\lb{s4}

For later convenience, we define the transfer function and the power spectrum of $\psi$ on the superhorizon scales as follows \cite{Inomata:2021zel},
\begin{equation}
\label{Tpsi}
T_{\psi}(k, \eta)=\frac{U_{\psi}(k, \eta)}{U_{\psi}(k, |k\eta|\ll1 )} ,
\end{equation}
\begin{equation}
\label{ps}
\mathcal{P}_{\psi}(k)= \frac{k^{3}}{2 \pi^{2}}\left|U_{\psi}(k, |k\eta|\ll1)\right|^{2}.
\end{equation}
Then the two-point function of $\psi$ can be expressed as
\begin{equation}
\left\langle\psi_{ \boldsymbol{k}}\left(\eta_{1}\right) \psi_{ \boldsymbol{k}^{\prime}}\left(\eta_{2}\right)\right\rangle=(2 \pi)^{3}  \delta\left(\boldsymbol{k}+\boldsymbol{k}^{\prime}\right) T_{\psi}\left(k, \eta_{1}\right) T_{\psi}^{\ast}\left(k, \eta_{2}\right) \frac{2 \pi^{2}}{k^{3}} \mathcal{P}_{\psi}(k).
\end{equation}
If the time derivative of a field is involved in the two-point function, the corresponding transfer function on the right-hand side of the above equation is just replaced by its time derivative.

The power spectra $\mathcal{P}^{A}_{h}(k,\eta)$ are related to the expectation values as
\begin{equation}
\label{hh}
   \left \langle h^A_{\bm{k}}(\eta) h^{A^{\prime}}_{\bm{k}'}(\eta)\right\rangle =(2\pi)^3\delta^{AA'}\delta^3(\bm k+\bm k')\frac{2\pi^2}{k^3}\mathcal{P}^{A}_{h}(k,\eta).
\end{equation}
Substituting eqs. \eqref{coz} and \eqref{GGreen} into Eq. \eqref{Eh}, we obtained the expectation values
\begin{equation}\label{ht}
\left\langle h_{\boldsymbol{k}}^{A}(\eta) h_{\boldsymbol{k}^{\prime}}^{A^{\prime}}(\eta)\right\rangle=\frac{16}{(z^A)^2} \int_{\eta_i}^{\eta} \mathrm{d} \eta_{1} \int_{\eta_i}^{\eta}  \mathrm{d} \eta_{2}~\frac{a(\eta_1)a(\eta_2)}{a(\eta)^2}G^A_{\bm k}\left(\eta, \eta_{1}\right) G^A_{{\bm k}^{\prime}}\left(\eta, \eta_{2}\right)\left\langle \mathcal S_{\boldsymbol{k}}^{A}\left(\eta_{1}\right) \mathcal S_{\boldsymbol{k}^{\prime}}^{A^{\prime}}\left(\eta_{2}\right)\right\rangle.
\end{equation}
After some lengthy but straightforward calculation, we obtain
\begin{align}
\label{ss}
\left\langle{\mathcal S}_{\boldsymbol{k}}^A\left(\eta_1\right) {\mathcal S}_{\boldsymbol{k}^{\prime}}^{A^{\prime}}\left(\eta_2\right)\right\rangle= & (2 \pi)^3 \delta\left(\boldsymbol{k}+\boldsymbol{k}^{\prime}\right) \delta^{A A^{\prime}} \frac{2 \pi^2}{k^3} \times\frac{k^4}{4}\Bigg\{\int_0^{\infty} \mathrm{d} v \int_{|1-v|}^{|1+v|} \mathrm{d} u\left[\frac{4 v^2-\left(1+v^2-u^2\right)^2}{4 u v}\right]^2 \nb\\
& \times  F\left(u , v , x_1\right)F^*\left(u , v , x_2\right) \mathcal{P}_{\psi}(u k) \mathcal{P}_{\psi}(v k)\Bigg\},
\end{align}
where $u\equiv|\bm k-\bm{p}|/k, v\equiv p/k$ and $x_1=k\eta_1$, $x_2=k\eta_2$, and
\begin{align}
F\left(u,v,x\right)=F_{\mathrm{GR}}\left(u,v,x\right)+F_{\mathrm{PV}}\left(u,v,x\right),
\end{align}
with
\begin{align}
\label{Fgr}
F_{\mathrm{GR}}(u,v, x)= &\frac{2}{\epsilon} T_{\delta\varphi}(ux)T_{\delta\varphi}(v x)+2T_{\psi}(ux)T_{\psi}(v x),
\end{align}
\begin{align}
\label{Fpv}
F_{\mathrm{PV}}(u,v, x)=&-\frac{\lambda^A k^2}{a^2 }\Bigg[\partial_{ x}\Big(f_{\varphi}\cdot \frac{2}{\sqrt{2\epsilon}\kappa\beta}T_{\delta \varphi}(u x)\cdot  T_{\psi}(v x)+f_{\varphi}\cdot  \frac{2}{\sqrt{2\epsilon}\kappa\beta}T_{\delta \varphi}(v x)\cdot  T_{\psi}(u x)\Big)\Bigg].
\end{align}
In deriving the eq. \eqref{ss}, we have used the relation
\begin{equation}
\left\langle\delta\varphi_{ \boldsymbol{k}}\left(\eta_{1}\right) \delta\varphi_{ \boldsymbol{k}^{\prime}}\left(\eta_{2}\right)\right\rangle=(2 \pi)^{3}  \delta\left(\boldsymbol{k}+\boldsymbol{k}^{\prime}\right) T_{\delta\varphi}\left(k, \eta_{1}\right) T_{\delta\varphi}^{*}\left(k, \eta_{2}\right) \frac{2 \pi^{2}}{k^{3}}\frac{2}{\epsilon\kappa^2\beta^2} \mathcal{P}_{\psi}(k),
\end{equation}
with
\begin{equation}
T_{\delta\varphi}(k, \eta)= \frac{U_{\delta\varphi}(k, \eta)}{U_{\delta\varphi}(k, |k\eta| \ll 1)},
\end{equation}
which is the transfer function of $\delta\varphi$.

Combining eqs. \eqref{hh}, \eqref{ht} and  \eqref{ss}, we can obtain the power spectra of SIGWs
\begin{align}
\label{ps-tensor}
\mathcal{P}^{A}_h(k,\eta)=\frac{4 }{\left(z^A(k)\right)^2}\int_{0}^\infty\mathrm{d}v\int_{|1-v|}^{1+v}\mathrm{d}u
\left[\frac{4v^2-(1+v^2-u^2)^2}{4uv}\right]^2| I^{A}(u,v,k,\eta)|^2\mathcal{P}_{\psi}(u k)\mathcal{P}_{\psi}(v k),\nb\\
\end{align}
where
\begin{align}
\label{IA}
I^{A}(u,v,k,\eta)
=&\int_{x_i}^x\mathrm{d}\bar{x}~k~\frac{a(\bar\eta)}{a(\eta)}G^A_{\bm k}(x,\bar{x})(F_{\mathrm{GR}}(u,v,\bar x)+F_{\mathrm{PV}}(u,v,\bar x))\nb\\
=&I_{\mathrm{GR}}(u,v,k,\eta)+I^{A}_{\mathrm{PV}}(u,v,k,\eta),
\end{align}
with $ x_i=k\eta_i$ satisfying $|x_i|\gg1$.

By employing the mode functions we obtain in the above section, the transfer functions for perturbations $\psi$ and $\delta\varphi$ are
\begin{equation}
\label{tf}
T_{\psi}(ux)=\mathrm{e} ^{-\mathrm{i}ux},\quad T_{\delta\varphi}(ux)=(1+\mathrm{i}ux )\mathrm{e} ^{-\mathrm{i}ux}.
\end{equation}
Substituting  eqs. \eqref{assume} and \eqref{tf} into eq. \eqref{IA}, we obtain
\begin{align}
\label{IGR}
I_{\mathrm{GR}}(u,v,x)
=&\frac{2}{\epsilon}\int_{x_i}^x\mathrm{d}\bar{x}~k~\frac{a(\bar\eta)}{a(\eta)}G^A_{\bm k}(x,\bar{x})\Big[(1+\mathrm{i}u\bar x)(1+\mathrm{i}v\bar x)\mathrm{e} ^{-\mathrm{i}(u+v)\bar x}+\epsilon\cdot\mathrm{e} ^{-\mathrm{i}(u+v)\bar x}  \Big]\nb\\
\simeq&\frac{2}{\epsilon} \int_{x_i}^x\mathrm{d}\bar{x}~k~\frac{a(\bar\eta)}{a(\eta)}G^A_{\bm k}(x,\bar{x})\Big[(1+\mathrm{i}u\bar x)(1+\mathrm{i}v\bar x)\mathrm{e} ^{-\mathrm{i}(u+v)\bar x} \Big],
\end{align}
\begin{align}
\label{IPV}
I^{A}_{\mathrm{PV}}(u,v,x)=&\frac{z_0k\lambda^A}{\epsilon}\int_{x_i}^x\mathrm{d}\bar{x}~k~\frac{a(\bar\eta)}{a(\eta)}G^A_{\bm k}(x,\bar{x}) \Bigg\{\Big[(1+\mathrm{i}u\bar x)\mathrm{e} ^{-\mathrm{i}(u+v)\bar x}+(1+\mathrm{i}v\bar x)\mathrm{e} ^{-\mathrm{i}(u+v)\bar x}\Big] \nb\\
&~~~~~~~~~-\bar x\Big[u^2\bar x\mathrm{e} ^{-\mathrm{i}(u+v)\bar x} -\mathrm{i} v(1+\mathrm{i}u\bar x)  \mathrm{e}^{-\mathrm{i} (u+v)\bar{x}}\Big] \nb\\
&~~~~~~~~~-\bar x\Big[v^2\bar x\mathrm{e} ^{-\mathrm{i}(u+v)\bar x} -\mathrm{i} u(1+\mathrm{i}v\bar x)  \mathrm{e}^{-\mathrm{i} (u+v)\bar{x}}\Big] \Bigg\}.
\end{align}
The explicit and lengthy expressions of $I_{\mathrm{GR}}$ and $I^{A}_{\mathrm{PV}}$ can be found in appendix \ref{app2}.

The magnitude of the parity violation in the GWs is conveniently characterized by the degree of the circular polarization, which is defined by \cite{Saito:2007kt,Gluscevic:2010vv}
\begin{equation}
\label{pi}
\Pi=\frac{\mathcal{P}^R_h-\mathcal{P}^L_h}{\mathcal{P}^R_h+\mathcal{P}^L_h}.
\end{equation}
From the expressions of $\mathcal{P}^{A}_h$ and $\Pi$, combing eqs. \eqref{IGR} and \eqref{IPV}, we find that the degree of circular polarization is large only if the contribution to SIGWs from PV term is about the same order as that from GR, $\mathcal{O}(I^A_{\mathrm{PV}})\sim \mathcal{O}(I_{\mathrm{GR}})$, i.e., $z_0k=k_{phys}/M_{\mathrm{PV}}\sim \mathcal{O}(1)$, otherwise, $|\Pi|\ll1$.

In order to avoid the appearance of ghost fields, the maximum wavenumber that we consider should satisfy $k^{max}_{phys}<M_{\mathrm{PV}}$. Note that $M_{\mathrm{PV}}$ is independent of $k_{phys}$, which means that $k_{phys}\ll M_{\mathrm{PV}}$ on large scales under the condition $k_{phys}\ll k^{max}_{phys}$. With this consideration, we have
\begin{equation}
z_0k=\frac{k_{phys}}{M_{\mathrm{PV}}}\ll 1,
\end{equation}
on large scales.

Based on the  above discussion, we can conclude that the contribution from the PV term is negligible, thus the degree of circular polarization of SIGWs generated during slow roll inflation is very small on large scales. As a result, it is difficult to test if gravitational interaction violates the parity in the early universe with SIGWs by CMB observation.

Up to now, in the main discussion of our paper, we have chosen an exponential form for the coupling function and  obtained the aforementioned results.
One may wonder the robustness of our conclusions based on such a particular choice of the couple function. Therefore, in the following, we will briefly discuss the SIGWs with a linear coupling function, $f(\varphi) = f_0\varphi$, which preserves the shift symmetry.

Combining the evolution of the background and the definition of $B^A$,
we can derive an expression given by:
\begin{equation}
\frac{B^{A''}}{B^A} \simeq \frac{a''}{a} + \frac{\lambda^A}{2}\frac{k_{\text{phys}}}{M_{\text{PV}}}\frac{a''}{a}.
\end{equation}
When  considering the SIGWs on large scales, we have $k_{\text{phys}}/M_{\rm{PV}}\ll 1$.
Consequently, the difference between Green's functions corresponding to exponential and linear coupling functions can be considered negligible.
On the other hand, the kernel $I^A_{\rm {PV}}$ can  be expressed as follows:
\begin{align}
&I^{A}_{\mathrm{PV}}(u,v,k,\eta)\nb\\
=&\frac{1}{\epsilon}\int_{x_i}^x\mathrm{d}\bar{x}~k~\frac{a(\bar\eta)}{a(\eta)}G^A_{\bm k}(x,\bar{x}) \Bigg\{\left(\frac{\lambda^A  k_{phys}}{M_{\mathrm{PV}}} \right) \Big[-\mathrm{i} (1+\mathrm{i}u \bar x+\mathrm{i}v \bar x)(u\bar x+v\bar x)e^{-\mathrm{i}(u \bar x+v\bar x)} \Big]\Bigg\}
\end{align}
where the integrand in the above equation contains the  coefficient $k_{phys}/M_{\mathrm{PV}} $, which suppresses the magnitude compared to $I^{A}_{\mathrm{GR}}$ \eqref{IGR}.
In conclusion, a linear form of coupling function doesn't alter our results, that is $\Pi \ll 1$ on large scales.

\section{Conclusion}\lb{s5}

In this paper, we made an analysis of the SIGWs from Chern-Simon gravity during slow roll inflation in detail. With slow roll approximation, we obtain the solution of scalar field $\varphi$ and the mode functions of perturbations $\delta\varphi$ and $\psi$. With the solution of scalar field $\varphi$, by taking the exponential form of the coupling function, we obtained the analytic expression for the Green's function, which is used to solve the SIGWs. We define the transfer functions $T_{\delta\varphi}$ and $T_{\psi}$ with the mode functions of the perturbation of the scalar field $U_{\delta\varphi}$ and metric perturbation $U_{\psi}$. Combining these transfer functions and the Green's function, we obtained a semianalytic expression for the power spectra $\mathcal{P}^A_{h}$ of SIGWs.

As shown in eqs. \eqref{IA}-\eqref{IPV} and \eqref{B-IGR}-\eqref{B-IA},
 the magnitude of the PV term $I^A_{\mathrm{PV}}$ is nearly $\mathcal{O}\left(z_0kI_{\mathrm{GR}}\right)$, where $z_0k=k_{phys}/M_{\mathrm{PV}}\ll1$ on large scales with  $k_{phys}\ll k^{max}_{phys}$. Obviously, the contribution from $I^A_{\mathrm{PV}}$ to the power spectra $\mathcal{P}^A_{h}$ is negligible, which results in a small degree of circular polarization of SIGWs. Furthermore,  we conducted a brief analysis of the case with a linear coupling function. Our analysis indicates that the above conclusion based on an exponential coupling function result is not altered and thus robust.

Our results show that on large scales, it is difficult to probe the parity violation in the CS gravity with the SIGWs generated during slow roll inflation.
It is therefore interesting to investigate the SIGWs in more general PV gravity theories, which may enjoy PV features that can be probed by the future observations.

\begin{acknowledgments}
This work was partly supported by the National Natural Science Foundation of China (NSFC) under the grant No. 11975020 and No. 12005309.
The equations of motion for the SIGWs from CS gravity are derived with the help of the Mathematica package \textit{xPand} \cite{Pitrou:2013hga}.
\end{acknowledgments}

\appendix

\section{The equations of motion}\label{app1}

In this appendix, we will show the equations of motion for the background and first-order scalar perturbations.

At the zeroth order, the background equations are
\begin{align}
\label{eom01}
3\mathcal{H}^2=&\kappa^{2}a^2\left(\frac{1}{2a^2} {\varphi^{\prime}}^{2}+V(\varphi)\right), \\
\label{eom02}
-\mathcal{H}^2-2\mathcal{H}^{\prime}=&\kappa^{2}a^2\left(\frac{1}{2a^2} {\varphi^{\prime}}^{2}-V(\varphi)\right).
\end{align}
The Klein-Gordon equation for the scalar field $\varphi$ is
\begin{equation}
\label{seq0}
\varphi^{\prime \prime}+2\mathcal{H}\varphi^{\prime}=-a^{2}~V_{\varphi}.
\end{equation}
In order to solve the SIGWs, we need the equations of motion of the first-order scalar perturbations, which are as follows
\begin{align}
\label{eom100}
\nabla^2\psi-3\mathcal{H}(\psi'+\mathcal{H}\phi) =&\frac12\kappa^{2}(\varphi^{\prime}{\delta \varphi}'-{\varphi^{\prime}}^2\psi+a^{2}~V_{\varphi}\delta \varphi),\\
\label{eom10i}
   \psi'+\mathcal{H}\phi=&\frac12 \kappa^{2}\varphi^{\prime}\delta \varphi,\\
\label{eom1ij}
\psi''+2\mathcal{H}\psi'+\mathcal{H}\phi'+(\mathcal{H}^2+2\mathcal{H}')\phi=&\frac12\kappa^{2}\Big(\varphi^{\prime}{\delta \varphi}'-{\varphi^{\prime}}^2{\psi}-a^{2}~V_{\varphi}\delta \varphi\Big),\\
\lb{phi-psi}
\psi-\phi=&0,
\end{align}
and
\begin{equation}
\label{seq1}
 \delta \varphi^{\prime \prime}+2\mathcal{H}\delta \varphi^{\prime}-\nabla^2 \delta \varphi-4 \psi^{\prime} \varphi^{\prime}=-a^{2}\delta \varphi V_{\varphi\varphi}-2\psi V_{\varphi}.
\end{equation}

Note that the PV term does not contribute to the equations of motion for the background evolution and the linear scalar perturbations.

\section{The analytic expression of  $I^{A}$  }\lb{app2}

In the limit $x\rightarrow 0$, the kernels defined in eqs. \eqref{IGR}-\eqref{IPV} become
\begin{equation}
I_{\mathrm{GR}}=\mathcal{I}_{\mathrm{GR}}(u,v,x\rightarrow 0)- \mathcal{I}_{\mathrm{GR}}(u,v,x_i),\quad
I^A_{\mathrm{PV}}=\mathcal{I}^A_{\mathrm{PV}}(u,v,x\rightarrow 0)- \mathcal{I}^A_{\mathrm{PV}}(u,v,x_i),
\end{equation}
where ~$\mathcal{I}_{\mathrm{GR}}$ and $\mathcal{I}^A_{\mathrm{PV}}$ are defined by
\begin{equation}
\mathcal{I}_{\mathrm{GR}}(u,v,y)=\frac{2}{\epsilon}
\int \mathrm{d} y ~\left[\frac{y \cos (y)-\sin (y)}{y^2} \right] \cdot\left[\left(1+\mathrm{i} uy\right)\left(1+\mathrm{i} vy\right)  \mathrm{e}^{-\mathrm{i} (u+v)y}\right],
\end{equation}
and
\begin{align}
\mathcal{I}^A_{\mathrm{PV}}(u,v,y)=&\frac{z_0k\lambda^A}{\epsilon}\int \mathrm{d} y ~\left[\frac{y \cos (y)-\sin (y)}{y^2} \right] \cdot \Bigg\{\Big[(1+\mathrm{i}uy)\mathrm{e} ^{-\mathrm{i}(u+v)y}+(1+\mathrm{i}vy)\mathrm{e} ^{-\mathrm{i}(u+v)y}\Big] \nb\\
&~~~~~~~~~-y\Big[u^2y~\mathrm{e} ^{-\mathrm{i}(u+v)y} -\mathrm{i} v(1+\mathrm{i}uy)\mathrm{e}^{-\mathrm{i} (u+v)y}\Big] \nb\\
&~~~~~~~~~-y\Big[v^2y~\mathrm{e} ^{-\mathrm{i}(u+v)y} -\mathrm{i} u(1+\mathrm{i}vy)  \mathrm{e}^{-\mathrm{i} (u+v)y}\Big] \Bigg\},
\end{align}
respectively.
After tedious manipulations, the concrete expressions of $\mathcal{I}_{\mathrm{GR}}$ and $\mathcal{I}_{\mathrm{PV}}$ are found to be
{\footnotesize
\begin{align}
&\mathcal{I}_{\mathrm{GR}}(u,v,y)\nb\\
=&\frac{1}{\epsilon}\Bigg\{-(u+v)\left(\frac{\cos[(u+v+1)y]}{u+v+1}+\frac{\cos[(u+v-1)y]}{u+v-1}\right)+uv\left(\frac{\cos[(u+v-1)y]}{u+v-1}-\frac{\cos[(u+v+1)y]}{u+v+1}\right)\nb\\
&+\mathrm{i}(u+v)\left(\frac{\sin[(u+v-1)y]}{u+v-1}+\frac{\sin[(u+v+1)y]}{u+v+1}\right)-\mathrm{i}uv\left(\frac{\sin[(u+v-1)y]}{u+v-1}-\frac{\sin[(u+v+1)y]}{u+v+1}\right)\nb\\
&-uv\left(\frac{y\sin[(u+v-1)y]}{u+v-1}+\frac{\cos[(u+v-1)y]}{(u+v-1)^2}+\frac{y\sin[(u+v+1)y]}{u+v+1}+\frac{\cos[(u+v+1)y]}{(u+v+1)^2}\right)\nb\\
&+\mathrm{i}uv\left(-\frac{y\cos[(u+v+1)y]}{u+v+1}+\frac{\sin[(u+v+1)y]}{(u+v+1)^2}-\frac{y\cos[(u+v-1)y]}{u+v-1}+\frac{\sin[(u+v-1)y]}{(u+v-1)^2}\right)\nb\\
&+\left(-\frac{\sin[(u+v-1)y]}{y}+\frac{\sin[(u+v+1)y]}{y}\right)+\mathrm{i}\left(-\frac{\cos[(u+v-1)y]}{y}+\frac{\cos[(u+v+1)y]}{y}\right)\Bigg\},
\end{align}
}
and
{\footnotesize
\begin{align}
&\mathcal{I}^A_{\mathrm{PV}}(u,v,y)\nb\\
=&\frac{z_0k\lambda^A}{\epsilon}
\Bigg\{- (u+v) \left(\frac{\cos [(u+v-1)y]}{ (u+v-1)}+\frac{\cos [ (u+v+1)y]}{ (u+v+1)}+ \frac{\mathrm{i}\sin [(u+v-1)y] }{ (u+v-1)}+\frac{\mathrm{i}\sin [ (u+v+1)y]}{ (u+v+1)}\right)\nb\\
&+\frac{1}{2}(u+v)^2 \left(\frac{\cos [(u+v-1)y]}{ (u+v-1)}-\frac{\cos [ (u+v+1)y]}{ (u+v+1)}- \frac{\mathrm{i}\sin [(u+v-1)y]}{ (u+v-1)}+\frac{\mathrm{i}\sin [ (u+v+1)y]}{ (u+v+1)}\right)\nb\\
&-\frac{1}{2} (u+v)^2 \left(\frac{\cos [(u+v-1)y]}{(u+v-1)^2}+\frac{\cos[ (u+v+1)y]}{(u+v+1)^2}+\frac{y \sin [(u+v-1)y]}{u+v-1}+\frac{y \sin[ (u+v+1)y]}{u+v+1}\right)\nb\\
&+\frac{1}{2} \mathrm{i} (u+v)^2 \left(\frac{\sin[(u+v-1)y]}{(u+v-1)^2}+\frac{\sin [ (u+v+1)y]}{(u+v+1)^2}-\frac{y \cos[(u+v-1)y]}{u+v-1}-\frac{y \cos  [ (u+v+1)y]}{u+v+1}\right)\nb\\
&-\left(\frac{\sin[(u + v - 1) y]}{y}-\frac{\sin[(u + v + 1) y]}{y}\right)-\mathrm{i}\left(\frac{\cos[(u + v - 1) y] }{y}-\frac{ \cos[(u + v + 1) y]}{y}\right)\Bigg\}.
\end{align}
}

With the above expressions, we can obtain $\mathcal{I}_{\mathrm{GR}}$ and $\mathcal{I}^A_{\mathrm{PV}}$ in the limit $y\rightarrow0$, which are
\begin{equation}
\mathcal{I}_{\mathrm{GR}}(u,v,y)|_{y\rightarrow0}=-\frac{1}{\epsilon}\frac{2(u^2+4 u v+v^2-1)}{(u+v-1)^2 (u+v+1)^2},~~~~~~
\end{equation}
and
\begin{equation}
\mathcal{I}^A_{\mathrm{PV}}(u,v,y)|_{y\rightarrow0}=-\frac{z_0k\lambda^A}{\epsilon}\frac{2 \left(2 u^2+4 u v+2 v^2-1\right)}{(u+v-1)^2 (u+v+1)^2},
\end{equation}
respectively.
Thus the kernels $I_{\mathrm{GR}}$ and $I^{A}_{\mathrm{PV}}$ in the limit $x\rightarrow0$ are
\begin{equation}
\label{B-IGR}
I_{\mathrm{GR}}(u,v,x\rightarrow0)=-\frac{1}{\epsilon}\frac{2(u^2+4 u v+v^2-1)}{(u+v-1)^2 (u+v+1)^2}- \mathcal{I}_{\mathrm{GR}}(u,v,x_i),~~~~~~
\end{equation}
and
\begin{equation}
\label{B-IPV}
I^A_{\mathrm{PV}}(u,v,x\rightarrow0)=-\frac{z_0k\lambda^A}{\epsilon}\frac{2 \left(2 u^2+4 u v+2 v^2-1\right)}{(u+v-1)^2 (u+v+1)^2}-\mathcal{I}^A_{\mathrm{PV}}(u,v,x_i),
\end{equation}
respectively.
Finally, we get the analytic expression of $I^{A}$ in the limit $x\rightarrow0$,
\begin{equation}
\label{B-IA}
| I^{A}(u,v,x\rightarrow0)|^2
=\Big|I_{\mathrm{GR}}(u,v,x\rightarrow0)+I^{A}_{\mathrm{PV}}(u,v,x\rightarrow0)\Big|^2.
\end{equation}

%

\end{document}